\documentclass[usenatbib]{mnras}
\usepackage{epsfig}
\usepackage{amsmath,amssymb}
\usepackage{color}
\usepackage{booktabs}
\usepackage{pdflscape}

\usepackage{epsfig}
\usepackage{color}
\usepackage{soul}
\usepackage{aas_macros}  
\usepackage{chngcntr}
\graphicspath{{figures/}}

  \newcommand{\Teff}{\mbox{\,\em T$_{\rm eff}$}}         
 \newcommand{\teff}{\mbox{\,$T_{\rm eff}$}}      
\newcommand{\lgcs}{\mbox{\,$\log g / {\rm cm\,s^{-2}}$}}        

  \newcommand{\nHe}{\mbox{\,$n_{\rm He}$}}               
  \newcommand{\kmsec}{\,\mbox{$\mbox{km}\,\mbox{s}^{-1}$}}    
  \def\simge{\mathrel{\raise1.16pt\hbox{$>$}\kern-7.0pt
    \lower3.06pt\hbox{{$\scriptstyle \sim$}}}}           
  \def\simle{\mathrel{\raise1.16pt\hbox{$<$}\kern-7.0pt
    \lower3.06pt\hbox{{$\scriptstyle \sim$}}}}           
\newcommand{\ledz}{UVO\,0825+15}   
\newcommand{\crimson}{LS\,IV$-14^{\circ}116$}  

\newcommand{\rooster}{HE\,2359--2844}  
\newcommand{\kraft}{HE\,1256--2738}  
\newcommand{\pbsalt}{EC\,22536-5304} 

\title[EC 22536-5304: a new lead-rich subdwarf]{EC 22536-5304: SALT identifies a new lead-rich intermediate helium subdwarf\thanks{Based on observations made with the Southern Africa Large Telescope (SALT) under programmes 2016-2-SCI-008, 2018-1-SCI-038  and 2018-2-SCI-033 (PI: Jeffery)}
}
\author[C. S.~Jeffery \& B. Miszalski]{C. S. Jeffery$^{1,2,3}$\thanks{email: simon.jeffery@armagh.ac.uk}, and B. Miszalski$^{4,5}$ \\
$^{1}$Armagh Observatory and Planetarium, College Hill, Armagh BT61 9DG, UK\\
$^{2}$School of Physics, Trinity College Dublin, College Green, Dublin 2, Ireland\\
$^{3}$Institute of Astronomy, University of Cambridge, Cambridge CB3 0HA, UK\\
$^{4}$South African Astronomical Observatory, PO Box 9, Observatory, 7935, South Africa\\
$^{5}$Southern African Large Telescope Foundation, PO Box 9, Observatory, 7935, South Africa\\
}

\begin{document}

\date{Accepted \ldots. Received \ldots; in original form \ldots}

\pagerange{\pageref{firstpage}--\pageref{lastpage}} \pubyear{2014}

\maketitle

\label{firstpage}

\begin{abstract}
SALT spectra of the helium-rich hot subdwarf \pbsalt\  show strong absorption lines of triply-ionized lead.
Analysis of the HRS spectrum and a follow-up SALT/RSS spectrum show \pbsalt\ to have surface properties (temperature, gravity, helium/hydrogen ratio) similar to other heavy-metal subdwarfs. With a lead overabundance of 4.8\,dex relative to solar, \pbsalt\ is the most lead-rich intermediate helium  subdwarf discovered so far.   
\end{abstract}


\section{Introduction}
\label{s:intro}

Hot subluminous stars generally have surfaces which are either extremely hydrogen-rich or helium-rich. 
The former are probably due to chemical diffusion in the stellar atmosphere which encourages the heavier helium to sink \citep{heber86}.  
The latter are probably due to a previous evolution which has involved the destruction of hydrogen \citep{iben90,han02,zhang12a}. 
A small group of hot subdwarfs with intermediate helium abundances has been identified \citep[e.g.][]{schulz91}, of which several show remarkable overabundances of certain iron-group and/or trans-iron elements \citep{edelmann03,naslim11,naslim13,jeffery17a,wild18}. 
The frequency with which these stars occur, the range of surface properties, and the reason for their extraordinary surface characteristics represent significant questions for subdwarf astronomy. 
This paper reports the analysis of an intermediate helium-rich  
hot subdwarf observed during a medium and high-resolution spectroscopic survey of southern hot subdwarfs with the Southern Africa Large Telescope (SALT; \citet{buckley06,odonoghue06}).


\section[]{Observations}
\label{s:obs}

\pbsalt\ (=GALEX J225635.7-524837.1, $\alpha_{2000} = 22^{\circ} 56\arcmin 35.76\arcsec$, $\delta_{2000} = -52^{\circ} 48\arcmin 37.71\arcsec$) was identified as a hot subdwarf in the Edinburgh Cape survey (Zone 4) \citep{kilkenny16}, and as a helium-rich sdOB star by \citet{geier17}. 

As part of an ongoing programme to identify and analyse early-type hydrogen-deficient stars of interest, 
\pbsalt\ was observed with the Southern Africa Large telescope (SALT) High Resolution Spectrograph (HRS) in medium resolution mode on 2017 May 18  and on 2018 November 15.  
Exposure times were $2\times2000$\,s on both occasions. 
The nominal instrumental resolution $R\approx 43\,000$. 
The mean signal-to-noise ratio in the continuum around $\lambda\lambda 4560 - 4580$\AA\ is $\approx17$ ($\sigma = 0.06$). 
The 2017 observation was reduced to wavelength-calibrated orders using the pyHRS pipeline \citep{crawford15,crawford16}, sky-subtracted and order-merged using bespoke echelle-management software. 
The pyHRS pipeline was no longer maintained by SALT at the time of the 2018 observations.  
Both 2017 and 2018 observations were reduced in an analogous way using the {\sc MIDAS} based pipeline \citep{Kniazev16,Kniazev17}. 
The MIDAS pipeline delivers a wavelength calibration robust to $<300$\,m\,s$^{-1}$, but has poorer signal-to-noise and blaze correction characteristics. 
Therefore we used the MIDAS-reduced product for radial velocities and the pyHRS-reduced product for all other purposes. 

A notable feature in the spectrum of \pbsalt\ is the presence of a strong isolated aborption line at $\lambda 4496$\AA, a wavelength corresponding to that of triply ionized lead (Pb{\sc iv} $\lambda 4496$\AA). 
Since this would indicate a star with comparable properties to the heavy-metal subdwarfs, it became a priority for follow-up analysis. 
This is the only Pb{\sc iv} line lying above the steep drop in HRS sensitivity around 4300\AA.  

With relatively short wavelength ranges covered by individual \'echelle orders,  \'echelle spectra are notoriously difficult to rectify in the  presence of broad hydrogen and helium absorption lines. 
To provide confirmation,  \pbsalt\ was observed  with the SALT Robert Stobie Spectrograph (RSS; \cite{burgh03,kobulnicky03}) using grating PG2300 on 2018 June 4. 
The RSS detector consists of three adjacent charge-coupled devices, separated by two gaps. 
Double exposures were taken at two different grating angles to provide a continuous spectrum in the wavelength range 3850 -- 5150 \AA\, and to more easily reject cosmic-ray contamination. 
Exposure times were $2\times50$\,s at camera angle 63.95$^{\circ}$ and $2\times70$\,s at 60.05$^{\circ}$.
The mean instrumental resolution  $R\approx 3600$.
The mean signal-to-noise ratio in the continuum around $\lambda\lambda 4560 - 4580$\AA\ is $\approx100$ ($\sigma = 0.01$). 
Basic data processing was applied using the \textsc{pysalt}\footnote{http://pysalt.salt.ac.za} package \citep{crawford10} and then reduced using standard \textsc{iraf} tasks and the \textsc{lacosmic} package \citep{dokkum01} as per the process described by \citet{koen17}. Contemporaneous arc lamp exposures were used to determine the wavelength solution for each spectrum.
The one-dimensional spectra were extracted using the \textsc{apall} task and these were rectified using low-order polynomials fitted to regions of continuum identified autonomously, 
and merged  using weights based on the number of photons detected in each segment of spectrum used as input. 

\begin{figure*}
\epsfig{file=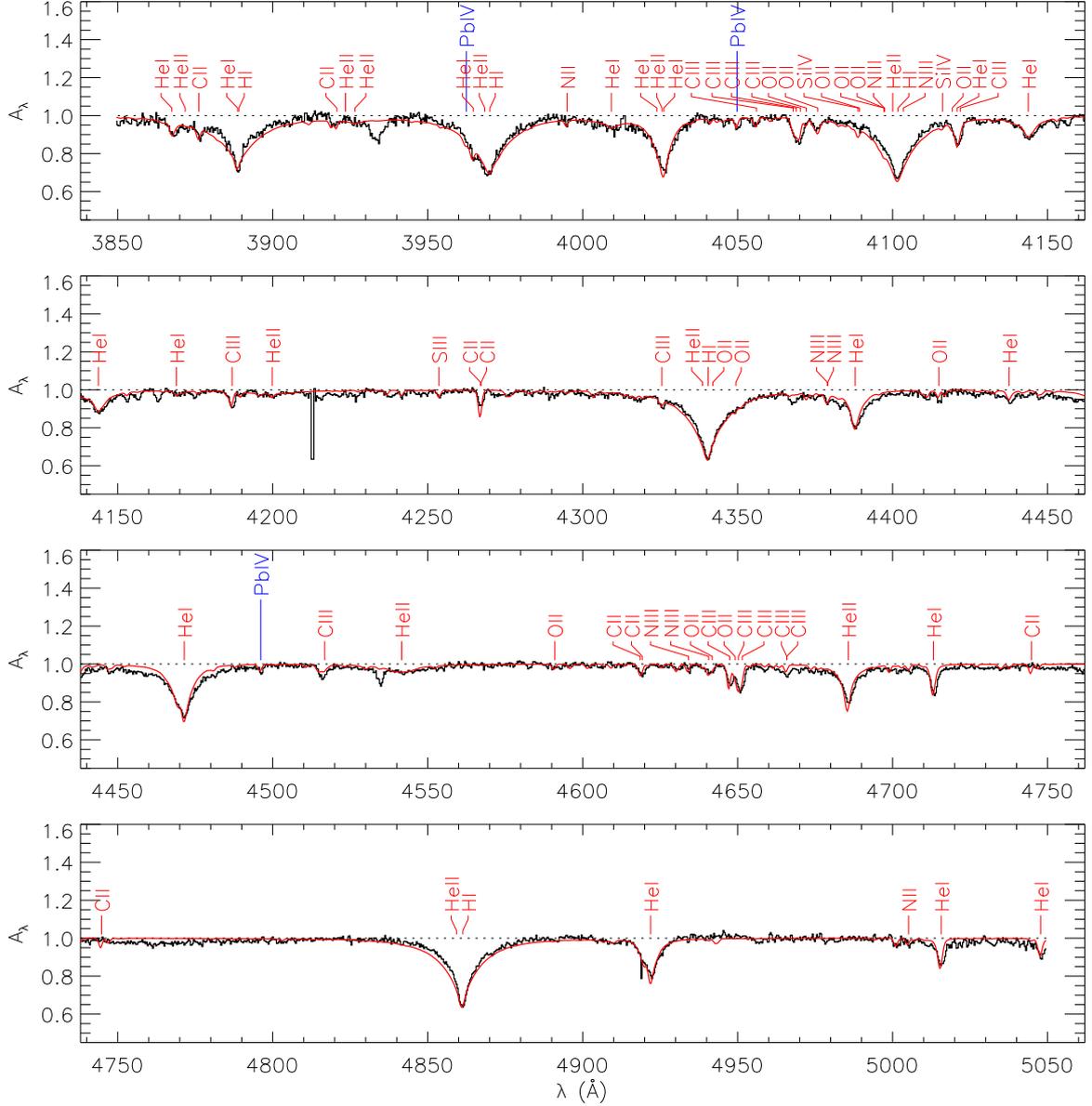,width=0.9\textwidth}
\caption{The observed SALT/RSS spectrum of \pbsalt\ (black histogram) 
and a model (red, grey in print) having  $\teff=36\,000$\,K, $\lgcs=5.75$,  $n_{\rm He}=0.20$, $v_{\rm turb}=5\kmsec$ 
and abundances shown in Table~\ref{t:abunds}.     
Lines which, if isolated, have theoretical equivalent widths greater than $40$m\AA\  are labelled. 
The model has been degraded to the instrumental resolution of 1.2\AA\ (FWHM). }
\label{f:atlasA}
\end{figure*}

\begin{figure*}
\epsfig{file=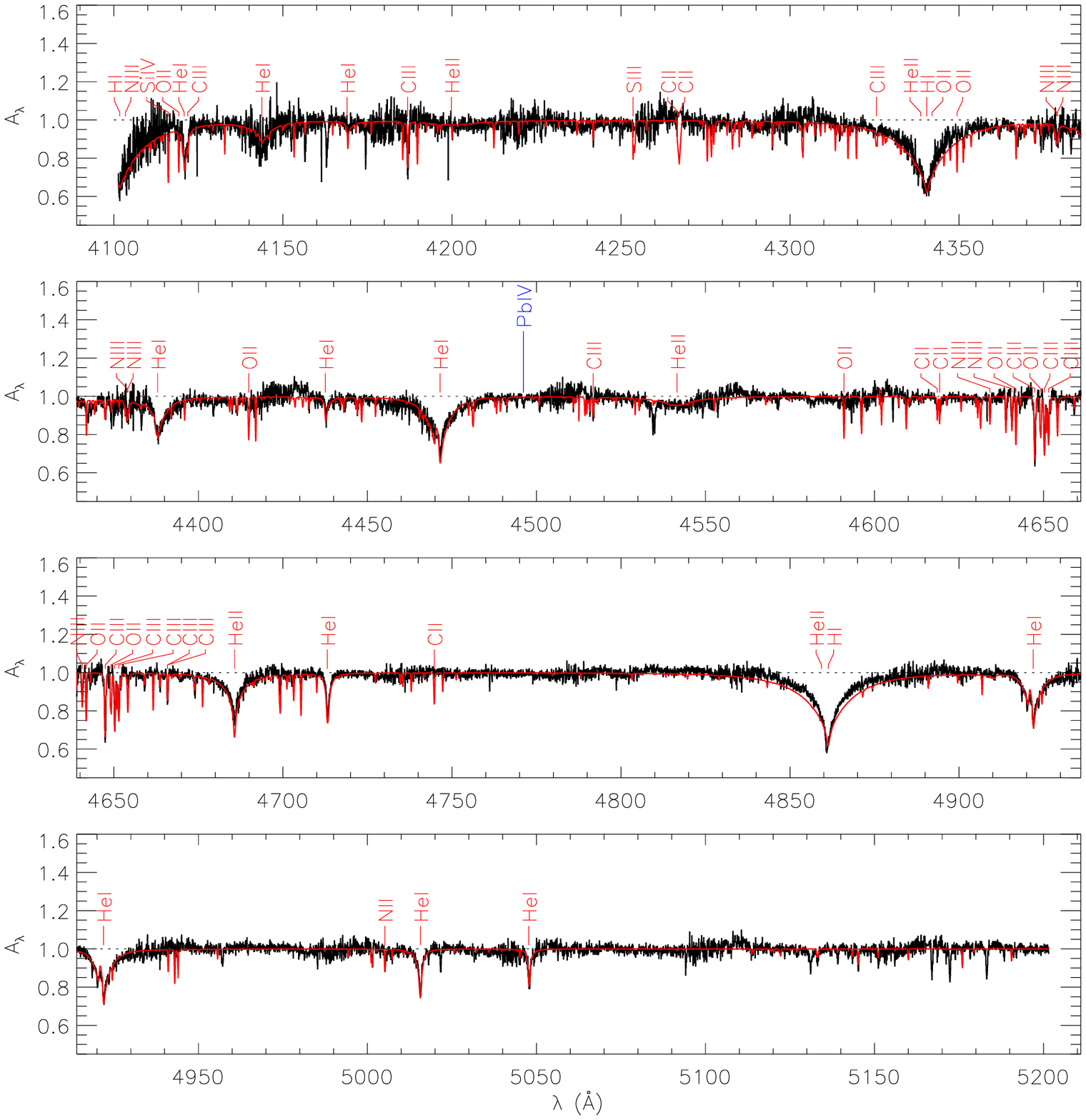,width=0.9\textwidth}
\caption{The SALT/HRS spectrum of \pbsalt\ (black histogram) and the best-fit (interpolated) model (B$_{\rm hrs}$: red, grey in print) having  $\teff=35\,560$\,K, $\lgcs=5.91$,  $n_{\rm He}=0.17$, $v_{\rm turb}=5\kmsec$  and abundances from the {\bf hiz} grid.     
The lines have been labelled using criteria from Fig.\,\ref{f:atlasA}. 
The observed spectrum and model have been resampled at a dispersion of 0.05\AA/pixel.
The model is convolved with an instrumental profile of 0.03\AA\ (FWHM).}
\label{f:atlasB}
\end{figure*}

The resulting spectra are shown in compact form in Figs.~\ref{f:atlasA} and \ref{f:atlasB}. 
Significantly, the absorption line at  $\lambda 4496$\AA\ is strong enough to appear in the RSS spectrum. 
There is also a stronger feature present at $\lambda 4050$\AA\ which corresponds (in wavelength and strength) to Pb{\sc iv} $\lambda 4050$\AA, confirming the identification of the longer wavelength line in the HRS spectrum. 

\begin{table}
\caption[Atmospheric parameters]
   {Atmospheric parameters for {\pbsalt}. Parentheses show formal errors in the last significant figure. Solutions B
   were obtained with additional weight being given to the principal helium lines. Formal errors in the last digits are in parentheses. }
\label{t:atmos}
\small
\setlength{\tabcolsep}{2pt}
\begin{center}
\begin{tabular}{l llll}
\hline
Solution      & A$_{\rm rss}$ &  B$_{\rm rss}$  & A$_{\rm hrs}$ & B$_{\rm hrs}$ \\
\hline
$\Teff$/kK   & $37.58(06)$ & $36.91(10)$ & $36.37(15)$ & $35.56(25)$ \\
$\lgcs$      &  $5.54(01)$ &  $6.11(02)$ &  $5.16(03)$  & $5.91(01)$ \\
$n_{\rm He}$ & $0.331(04)$ & $0.251(04)$ & $0.336(12)$  & $0.166(10) $ \\
$\log y$     & $-0.306(04)$ & $-0.475(08)$ & $-0.296(12)$  & $-0.701(40)$  \\ 
$v_{\rm rot} \sin i$/km\,s$^{-1}$ & -- & -- & $<5$ & $<5$ \\
\hline
\end{tabular}
\end{center}
\end{table}

\section[]{Analysis}
\label{s:fits}

\subsection{Radial velocity}
The radial velocity (RV) of \pbsalt\ was measured  by comparing the SALT/HRS spectrum with a theoretical spectrum 
for a helium-rich subdwarf. 
In this case, there are substantial differences in RV between the spectra reduced with pyHRS and with MIDAS; on the basis of substantial external testing \citep{kniazev16.hrs-rv}, we adopt the MIDAS RVs.  
On the dates observed the heliocentric velocities were: \\
\noindent $2017\,05\,18: ~~ -10.0 \pm 0.5 \kmsec $ and \\
\noindent $2018\,11\,18: ~~ +19.5 \pm 0.5 \kmsec. $\\
The above values and errors were obtained by comparing cross-correlation velocities for several different spectral regions between 4300 and 5400 \AA. 
Measurements on the two dates are so significantly different that we suspect the radial velocity to be variable. 
Additional observations will be required to explain the difference. 

 The absolute RSS radial-velocity calibration is not reliable because the light paths for comparison-lamp and starlight beams are not identical. No velocity is reported for this spectrum. 

%
%
\begin{table*}
\caption[Atmospheric abundances]
   {Atmospheric abundances of  {\pbsalt}, heavy-metal helium-rich subdwarfs with similar $\teff$ and $g$, and the Sun.
    Abundances are given as $\log \epsilon$, normalised to $\log \Sigma \mu \epsilon = 12.15$. Errors are given in parentheses. }
   \label{t:abunds}
\setlength{\tabcolsep}{3pt}
\begin{tabular}{@{\extracolsep{0pt}}p{22mm}l lll lll lll lll}
\hline
Star 					& H			& He 		& C			& N			& O			& Pb	\\
\hline        
\pbsalt                 & 11.90     & 11.20(05) & 8.74(23)  & 8.02(16)  & 8.24(15)  & 6.53(25)\\[1mm]
\crimson $^{1}$ 		& 11.83		& 11.23(20)	& 8.04(22)	& 8.02(20)	& 7.60(17)	& -- \\
\kraft$^{2}$	        & 11.45		& 11.44		& 8.90(54)	& 8.14(62)	& 8.08(10)	& 6.39(23)\\
\rooster$^{2}$       	& 11.58		& 11.38		& 8.51(29)	& 8.00(57)	& 7.81(16)	& 5.64(16) \\
\ledz$^3$				& 11.8  	& 11.2  	& $<6.5$	& 8.04(24)	& 7.43(07)	& 5.49(18) \\
PG\,1559+048$^4$        & 11.9      & 11.3      & 8.61(45)  & 8.02(21)  & $<7.0$    & 5.08(13) \\
FPS\,1749+373$^4$       & 11.8      & 11.3      & 8.52(27)  & 8.28(34)  & $<7.0$    & 4.96(11) \\
Sun$^{5}$				& 12.00		& 10.93		& 8.43		& 7.83		& 8.69		& 1.75 	 \\[3mm]
\hline
\end{tabular}\\
\parbox{170mm}{
References: 
1. \citet{naslim11}, 
2. \citet{naslim13},
3. \citet{jeffery17a},
4. \citet{naslim19},
5. \citet{asplund09}; photospheric except helium (helio-seismic). 
}
\end{table*}

\subsection{Photospheric parameters}

Analyses of both RSS and HRS  spectra were carried out using the Armagh LTE radiative transfer codes\footnote{Local thermodynamic equilibrium was assumed throughout the analysis} \citep{jeffery01b,behara06}. These include the model atmosphere code {\sc sterne}, the formal solution code {\sc spectrum} and the general purpose fitting package {\sc sfit}. 

For both spectra, effective temperature, surface gravity and surface helium and hydrogen abundances were obtained by finding the best fit spectrum in a grid of 
models computed with {\sc sterne} and {\sc spectrum}. 

The grid adopted covers a parameter space with  
\[
\begin{split} 
[\teff/{\rm kK},\lgcs,\nHe ] = \\
[ 26(04)42 , 3.8(0.4)6.0,  0.1(0.2)0.9,1.0 ].
\end{split}
\]
The first grid adopted ({\bf p00}) assumed a  solar distribution of metal abundances and a microturbulent velocity $v_{\rm turb}= 5\kmsec$ for both the calculation of
line opacities in the model atmosphere calculation (which affects the temperature stratification of the models) and for the formal solution, which affects relative line strengths and widths. 
In view of the highly non-standard abundances of helium-rich subdwarfs, it was  evident that several strong predicted lines (apart from H and He) are absent from the observed spectrum.
Since these could impact the quality of the fits, a second grid ({\bf hiz}) was computed  with reduced abundances of light elements, solar iron and enhanced calcium.
The grids are sampled over the wavelength interval  $4100 - 5200$\AA\ at intervals of 0.05\AA\ for analysing  HRS spectra, 
and over the wavelength interval  $3500 - 5500$\AA\ at intervals of 0.25\AA\ for analysing  RSS spectra. 

Best fits to the data were optimized by interpolating linearly in the model grid and using the Levenburg-Marquardt least-squares-minimisation option in {\sc sfit}. 
An essential ingredient in the fit procedure is the statistical error associated with each point in the observed spectrum. Here we used the values $\sigma$ reported in \S\,\ref{s:obs}. Whilst not ideal, at present none of the data reduction pipelines currently used for SALT spectroscopy preserves and propagates error estimates from raw data through to final product. In particular, the error associated with HRS \'echelle spectra varies considerably along each order.

A problem with fitting the blue-optical spectra of hot subdwarfs which have intermediate helium abundances is that it is difficult to fit both the hydrogen Balmer lines, the neutral helium lines and the ionized helium lines simultaneously. 
To illustrate, fits were first obtained including the full range of the spectra (3850 -- 5100\AA\ for RSS and 4100 -- 5200\AA\ for HRS) with all wavelengths equally weighted, 
The solutions are labelled A in Table\,\ref{t:atmos}, subscripted rss and hrs to indicate the spectrum measured.
The role of the principal helium lines in the fit was then enhanced by  reducing the relative weight of data points outside the vicinity of  He{\sc i} 4026, 4388, 4471 \AA\ and He{\sc ii} 4540, 4686\AA. The window sizes were $\pm10 - 25 $\AA\ depending on linewidth.
In practice, the error $\sigma$ associated with each data point outside these windows was increased by a factor 30.  
The resulting fit yielded the solutions labelled B in  Table\,\ref{t:atmos}.

Other factors which affect both the solution and the quality of the fit include:\\
-- the tolerance adopted for Levenburg-Marquardt convergence (0.0003). Increasing to 0.001 had no effect. \\
-- start values for the iteration. Adopted solutions were repeatable from different start values.\\
-- the model atmosphere grid. The {\bf hiz} grid was consistent with the final abundances reported in \S\,3.3.

The results for  RSS and  HRS spectra are consistent after allowing for the systematic differences in spectral normalisation, resolution and signal-to-noise ratio. 
The low surface gravity obtained in fit A$_{\rm hrs}$ reflects the different balance between H and He{\sc i} lines in the HRS and RSS spectra. 
It probably also reflects the difficulty of rectifying \'echelle spectral orders where these contain line profiles extend over nearly an entire \'echelle order or more. 

For this paper, we adopt the solution labelled B$_{\rm hrs}$ on the basis that the temperature difference between HRS and RSS solutions is smaller when the helium lines are given additional weight and that the HRS spectrum is more sensitive to subordinate line indicators of the ionization equilibrium. 
The small formal errors on $\log g$ shown in Table\,\ref{t:atmos} reflect the manner in which the Stark-broadened hydrogen and helium lines dominate the fit, so that once temperature is established, the gravity is well constrained. In practice the errors are not independent, and the error matrix should be investigated further.
The range of values shown in  Table~\ref{t:atmos} indicates the magnitude of systematic errors. 

Therefore we give the adopted surface properties for \pbsalt\ as  $T_{\rm eff} = 35\,550 \pm1\,500$\,K, $\lgcs = 5.92\pm 0.15$, and $\nHe=0.17\pm0.05$, where the errors include an allowance for systematic contributions.  
It is noted that this solution shown in Fig.~\ref{f:atlasB} gives the lowest helium abundance amongst those obtained.  

\begin{figure*}
       \epsfig{file=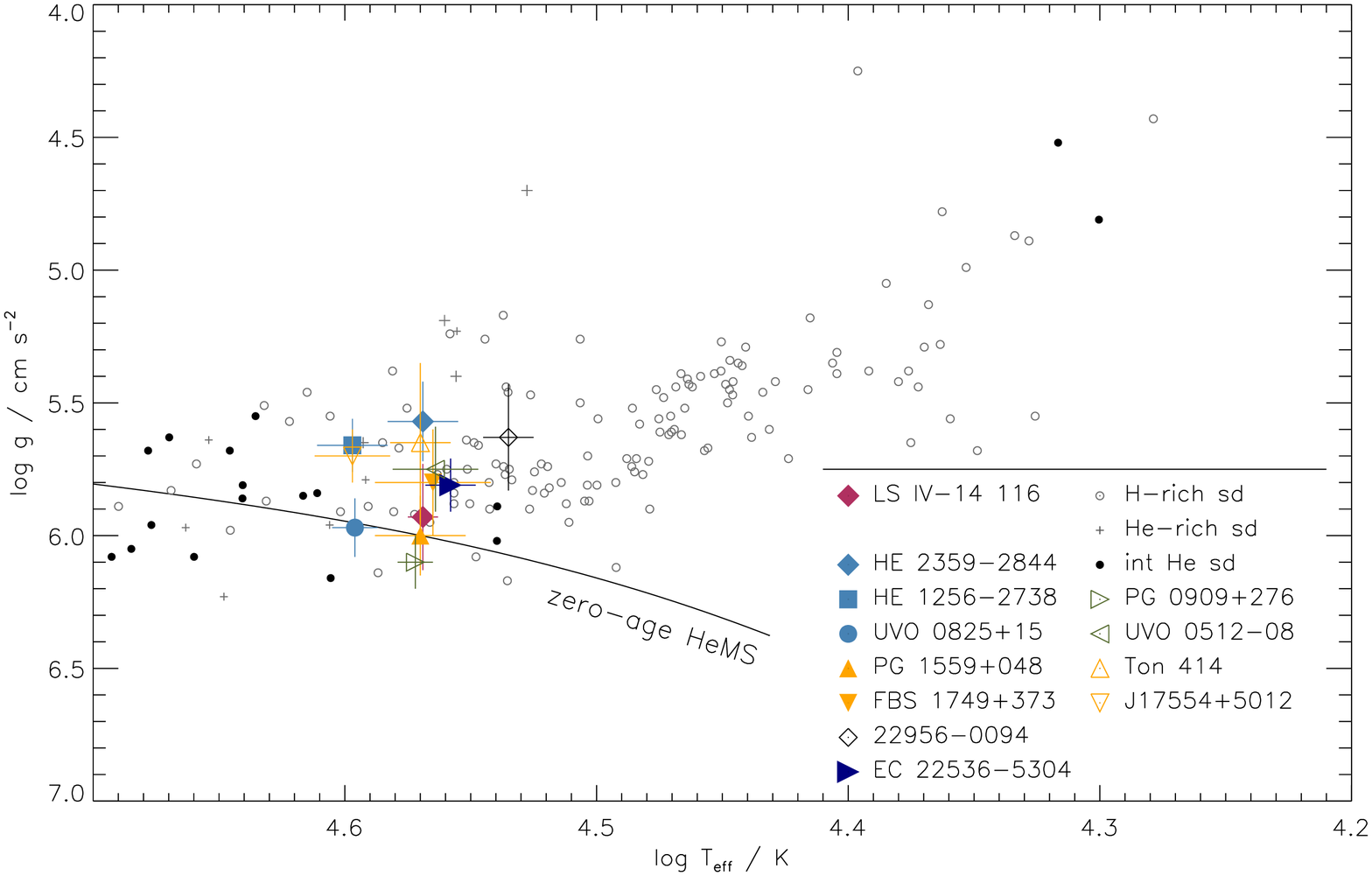,width=150mm}\\
       \epsfig{file=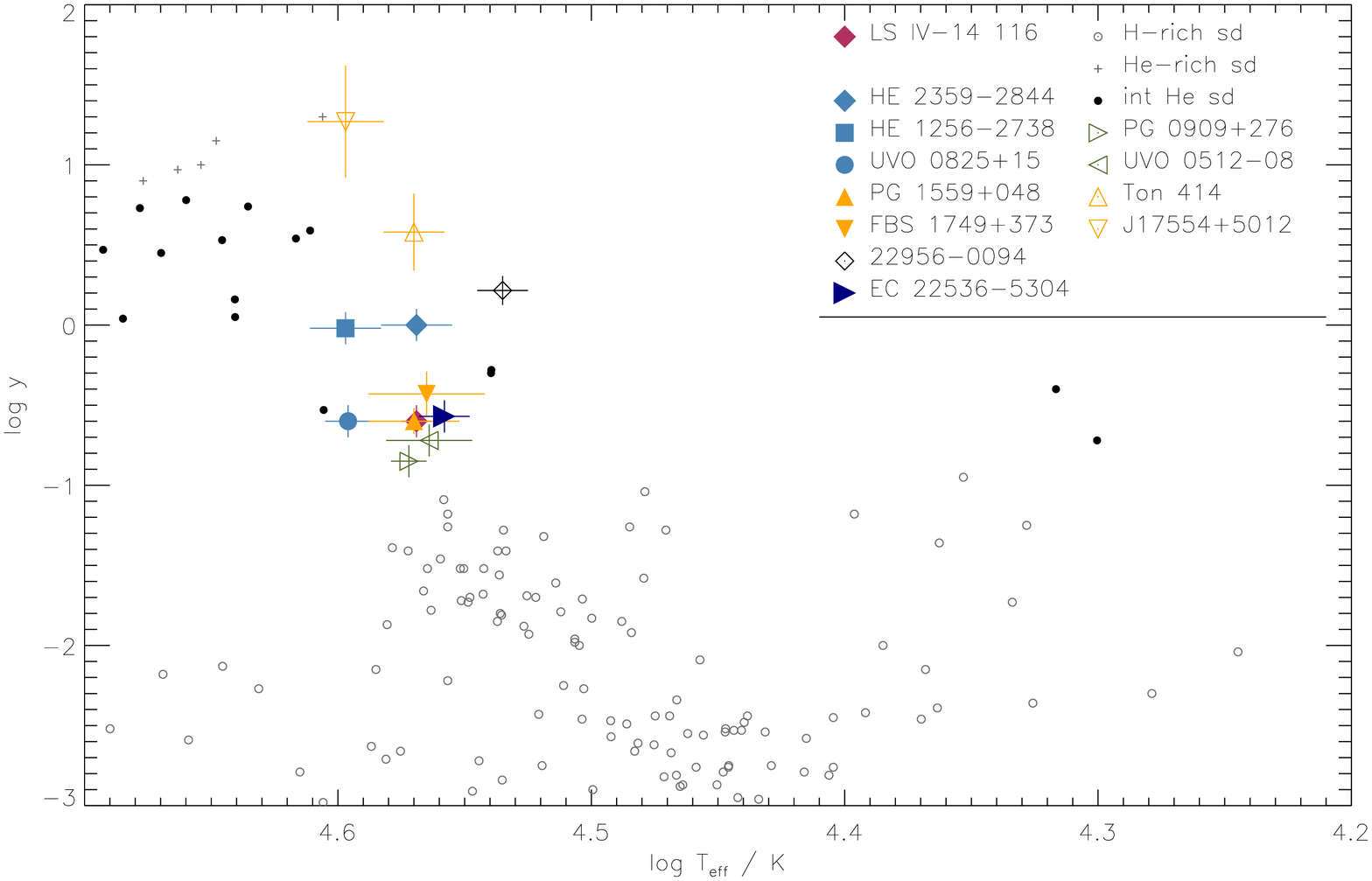,width=150mm}
        \caption{The distribution of \pbsalt\ and heavy-metal (large filled symbols), helium-rich and normal hot subdwarfs with effective temperature,  surface gravity (top) and helium abudance (bottom). The large open symbols are intermediate helium-rich subdwarfs which do not show an excess of heavy metals. The solid line shows a representative position for the theoretical zero-age helium main-sequence (HeMS: $Z=0.02$). The data are from this paper, \citet{naslim11,naslim13,naslim19,jeffery17a,nemeth12} and \protect\citet{wild18}.
} 
        \label{f:gt}
\end{figure*}

\subsection{Abundances}

Abundances of minor species may be obtained in a number of ways. 
In the line-by-line approach, equivalent widths of isolated or blended lines of the same species are used with {\sc spectrum} to solve directly for the line abundance using a Newton-Raphson iteration. 
In the spectrum-synthesis approach, the abundances of selected species can be optimised using {\sc sfit} which uses a Levenburg-Marquardt algorithm to minimize the mean square difference between fit and observed spectrum. 
The former is more direct but susceptible to the choice of lines -- a choice which can be used to avoid lines known to have poor atomic data or to be strongly affected by non-LTE, or which can omit the contributions of  weak lines which may be intrinsically more accurate by lying on the linear part of the curve-of-growth.
The latter can incorporate information from many weak lines, from the entire line profile, and can deal with multi-species blends.
After hydrogen and helium lines, the optical spectrum of \pbsalt\ is dominated by carbon, nitrogen and oxygen. 
With isolated exceptions, the relatively low signal-to-noise ratio in the HRS spectra precluded the clear identification of predicted lines of magnesium, aluminium, silicon, sulphur or iron. 
Attempts to infer abundances by optimisation other than for C, N and O were unsuccessful.  
In view of the noise, we executed a line-by-line analysis for the latter species and for the lead line at 4496\AA. 

The model atmosphere from which the formal solutions were calculated  
was chosen from the {\bf hiz} grid by proximity to solution B$_{\rm hrs}$, 
and had  $\teff/{\rm kK}=36$, {\bf $\lgcs=6.00$} and $\nHe=0.2$, and assumed $v_{\rm turb}=5\kmsec$. 

Equivalent widths were measured for 24 carbon lines, 26 nitrogen lines and 37 oxygen lines, as well as for \ion{Pb}{iv}4496\AA.
Abundances were derived for each line and averages formed for each ion. 
Eliminating lines which showed abundances different to the ion average by greater than 0.31 dex, overall average abundances 
were computed and are shown in Table\,\ref{t:abunds}. 
As an ionization equilibrium check on the measurement of \Teff, the abundance ratios of \ion{C}{ii/iii} and \ion{N}{ii/iii} were -0.12 and 0.01 dex respectively. 
Reducing the gravity to $\lgcs=5.75$ (solution A$_{\rm rss}$) changed the values in Table \,\ref{t:abunds} to $\log \epsilon_{\rm C, N, O, Pb} = 8.71(19), 8.01(14), 8.27(20), 6.53(25)$. 

 A regression of abundance versus equivalent width for  O{\sc ii} lines shows a clear positive gradient; efforts to reduce this gradient to zero usually involve increasing  $v_{\rm turb}$.  
This is ruled out by the core of the \ion{C}{\sc ii}4267\AA\ doublet which is resolved in the HRS observation  placing an upper limit of 5\kmsec\ to line broadening from all sources including rotation and turbulence \footnote{ 
The concept of microturbulence was originally introduced to account for abnormal broadening of hydrogen lines in the solar spectrum \citep{dejager54}, and subsequently as a diagnostic of convection in stellar photospheres. 
It is also used in a different sense, as here,  to explain why for a given abundance observed equivalent widths of strong lines are systematically weaker than predicted when compared with weak lines of the same species \citep{traving64,hundt73}. 
It is therefore a device used in the line profile calculation to increase overall line opacity preferentially for weak lines, rather than being a description of a physical process associated with line broadening. 
The requirement that atomic diffusion plays a key r\^ole in the chemical structure of hot subdwarf photospheres implies a dynamically  quiet environment. 
However this and the absence of observed broadening may not preclude the need for an opacity-enhancing device. 
Recent work on B stars suggests that a full non-LTE treatment and metallicity can play as strong a r\^ole as the microturbulent velocity \citep{dazsynska11}. }. 
Alternative explanations for the gradient include non-LTE effects and chemical stratification, neither of which can be addressed with the current data and models.  
Abundance analyses of normal subdwarfs frequently adopt $v_{\rm turb}=0$ \citep{heber00}. This would change the  
values in Table \,\ref{t:abunds} to $\log \epsilon_{\rm C, N, O, Pb} = 8.78(20), 8.05(13), 8.27(14), 6.63(25)$. 
The differences are $<0.1$\, dex and insubstantial. Nevertheless, the question of microturbulence should be investigated further with higher quality data for one or more of the heavy-metal subdwarfs. Such an investigation should also consider the consequences of extreme stratification due to selective radiative levitation in the stellar photosphere. 

An atlas of the spectrum showing the formal solution is shown in Fig.\,\ref{f:atlasA}.
Lines with theoretical equivalent widths $>40$m\AA\ (in the model) are labelled. 
Several observed lines are not present in the model. 
These lines are also visible in the spectrum of \ledz, and have not yet been identified \citep{jeffery17a}.

The abundance of lead measured from a single line in the HRS spectrum corresponds to nearly 5\,dex greater than solar. 
Both the 4496.2 and 4049.8\AA\ lines show up as weak features covering 2 pixels each in the RSS spectrum, with equivalent widths $35\pm20$ and $86\pm20$ m\AA (estimated errors). 
These correspond to abundances $\log \epsilon_{\rm Pb}= 6.36\pm0.30$ and $6.66\pm0.16$. 
The lead abundance in \pbsalt\ is therefore considerable; more precise data from high S/N high-resolution spectroscopy covering a wider spectral range is required.  

Since overabundances of other trans-iron elements are observed in intermediate helium-rich subdwarfs, it is useful to ask whether lead is the only such element with an elevated abundance in \pbsalt. 
By assigning an upper limit of 5 m\AA\ for the identification of lines in clean regions of the HRS spectrum, an upper limit $\log \epsilon_{\rm Zr} < 5.00$ is obtained from non-detection of  Zr{\sc iv}\,4569.3\AA, the strongest predicted zirconium line in the HRS spectral range.  This is 1.5 dex below the level observed in \crimson\ \citep{naslim10}, but does not exclude some enrichment above the solar value of 2.5. 


\section{Discussion}

Following the discovery of the first zirconium-rich intermediate-helium subdwarf and, subsequently, of similar lead-rich subdwarfs, an increasing number of heavy-metal subdwarfs have been discovered. 
Four have been published and a further five have been announced recently \citep{dorsch18,latour19a,naslim19}.
Figure \ref{f:gt} shows the distribution of published intermediate helium-rich subdwarfs in the $g-\Teff$   and $y-\Teff$ planes,  as well as the distributions of helium-rich and helium-poor subdwarfs. 
It includes helium-rich stars both with and without observed excesses of heavy elements. 
The former include \crimson\ \citep{naslim11}, HE\,2359--2844 and HE\,1256-2738 \citep{naslim13}, PG\,1559+048 and FBS\,1749+373 \citep{naslim19}.
The latter include BPS CS 22956-0094 \citep{naslim13}, PG\,0909+276, UVO\,0512-08 \citep{wild18}, Ton\,414 and GALEX\,J17554+5012 \citep{naslim19}. 
Amongst the second group, PG\,0909+276 and UVO\,0512--08 show high excesses of iron-group elements \citep{edelmann03th,wild18}. 
No extreme overabundances were detected in the remaining three, which are substantially more helium-rich. 
The figure illustrates how stars in approximately the same volume of $\Teff - g - y$ space show a considerable variety of enhanced chemistries. 

\pbsalt\ sits squarely amongst the heavy-metal subdwarfs, and possibly at the cool extremity of the current sample. 
It demonstrates how strongly concentrated selected species can become in the line forming region of the photosphere. 
Previous papers \citep{naslim11,green11,jeffery17a} have discussed the possible origin of the heavy-metal excess in terms of selective radiative levitation in the stellar photosphere. 
These ideas will be reviewed in the light of \pbsalt\ and other new discoveries in due course (Jeffery et al. in prep.). 

\section{Conclusion}

SALT HRS observations of the faint-blue He-sdOB star \pbsalt\ show it to be an intermediate helium-rich subdwarf. 
It shows very strong absorption lines due to triply ionized lead from which the inferred abundance inferred is some
4.8 dex above solar, making \pbsalt\ the most lead-rich subdwarf identified to date. 
For a hot subdwarf, the atmosphere is moderately rich in carbon and oxygen, with the abundances of both exceeding that of nitrogen. However, in this respect \pbsalt\ is no more exceptional than other intermediate helium-rich subdwarfs. 

Additional high-resolution observations are required to resolve the radial velocity behaviour of \pbsalt, and to provide sufficiently high-quality data to measure abundances of other species.  
Detailed model atmospheres including physics of radiative levitation and gravitational settling of these heavy metals need to be constructed to test the proposal that superabundances arise from chemical diffusion. 
Photometric observations repeated over several hours with a precision better than 0.5\% should be carried out to check for pulsations. 
Meanwhile, because of the exceptional strength of the signature absorption lines, these observations demonstrate the potential or SALT and RSS to discover additional heavy-metal subdwarfs. 

\section*{Acknowledgments}

The observations reported in this paper were obtained with the Southern African Large Telescope (SALT)
under programmes 2016-2-SCI-008, 2018-1-SCI-038  and 2018-2-SCI-033 (PI: Jeffery).
The authors are indebted to the hard work of the entire SALT team. 

CSJ thanks Churchill College Cambridge for a visiting by-fellowship and the Institute of Astronomy Cambridge for a visitor grant. 
The Armagh Observatory and Planetarium are funded by direct grant from the Northern Ireland Department for Communities. BM acknowledges support from the National Research Foundation (NRF) of South Africa. IRAF is distributed by the National Optical Astronomy Observatory, which is operated by the Association of Universities for Research in Astronomy (AURA) under a cooperative agreement with the National Science Foundation. 


\bibliographystyle{mn2e}
\bibliography{ehe}

\appendix
\renewcommand\thefigure{A.\arabic{figure}} 
\renewcommand\thetable{A.\arabic{table}} 
\section{Equivalent Width Measurements} 

Table\,\ref{t:eqwids} shows measured equivalent widths ($W_{\lambda}$) in m\AA\ and laboratory wavelengths for lines identified in the HRS spectrum of \pbsalt. 
Abundances derived from lines marked ":" were not included in the final means, having individual line abundances more than 0.31 dex from these values. 
 To save space, equivalent widths are rounded to whole numbers of m\AA\ and errors are shown in units of \AA$\times10^{-4}$, {\it i.e. an equivalent width of $6.2\pm0.1${\rm m\AA} is shown as $6\pm1$}. 
   The error estimate is primarily derived from the variance measured from continuum windows either side of the line being measured.

\begin{table}
\caption[Equivalent Widths]
   {Equivalent widths measured from the SALT/HRS spectrum of \pbsalt. Lines marked ":" were not included in the mean abundances.   }
\label{t:eqwids}
\small
\setlength{\tabcolsep}{3pt}

\begin{tabular}{ccc}
\begin{minipage}[!t]{26mm}
\begin{tabular}{lll}
Ion & \\
\multicolumn{3}{l}{$\lambda$ \hfill $W_{\lambda}$ \hfill $\delta W_{\lambda}$ } \\
\multicolumn{3}{l}{ \AA \hfill  m\AA \hfill $10^{-4}$\AA } \\

\hline
C{\sc ii}&\\
4313.1 & 6 &$\pm$1\\
4317.26 & 23: &$\pm$5\\
4372.35 & 36 &$\pm$6\\
4374.27 & 28 &$\pm$5\\
4411.2 & 19: &$\pm$4\\
4735.46 & 9: &$\pm$4\\
5132.95 & 32 &$\pm$3\\
5143.49 & 26 &$\pm$3\\
5145.16 & 40 &$\pm$4\\
5151.09 & 30: &$\pm$3\\[1mm]
C{\sc iii}&\\
4121.84 & 15: &$\pm$231\\
4186.9 & 149 &$\pm$208\\
4325.56 & 96: &$\pm$38\\
4515.78 & 32 &$\pm$16\\
4516.77 & 48 &$\pm$29\\
4647.42 & 179 &$\pm$508\\
4650.25 & 136: &$\pm$281\\
4651.47 & 92 &$\pm$122\\
4652.06 & 31 &$\pm$28\\
4659.06 & 30 &$\pm$5\\
4663.64 & 37 &$\pm$10\\
4665.86 & 65 &$\pm$15\\
4673.95 & 45: &$\pm$4\\[1mm]
4171.6 & 14 &$\pm$3\\
4176.16 & 19 &$\pm$3\\
4199.98 & 21: &$\pm$5\\
\end{tabular}
\end{minipage}
&
\begin{minipage}[!t]{26mm}
\begin{tabular}{lll}
\multicolumn{3}{l}{N{\sc ii}~(contd.)}\\
4227.74 & 14 &$\pm$1\\
4241.18 & 28 &$\pm$4\\
4432.74 & 25 &$\pm$14\\
4447.03 & 29 &$\pm$2\\
4530.40 & 25 &$\pm$9\\
4607.16 & 17 &$\pm$4\\
4630.54 & 18: &$\pm$1\\
4643.09 & 27: &$\pm$18\\
5001.13 & 48 &$\pm$8\\
5005.15 & 38 &$\pm$5\\
5007.33 & 9: &$\pm$1\\
5025.66 & 12: &$\pm$1\\
5045.09 & 8 &$\pm$1\\[1mm]
N{\sc iii}&\\
4200.10 & 21 &$\pm$6\\
4195.76 & 19 &$\pm$3\\
4200.10 & 17 &$\pm$4\\
4510.91 & 16 &$\pm$5\\
4514.86 & 31 &$\pm$11\\
4518.15 & 11 &$\pm$3\\
4523.58 & 18 &$\pm$2\\
4634.14 & 35: &$\pm$5\\
4640.64 & 71 &$\pm$30\\
4641.85 & 59: &$\pm$19\\[1mm]
O{\sc ii}&\\
4119.22 & 46 &$\pm$23\\
4132.80 & 19 &$\pm$2\\
4169.22 & 18: &$\pm$1\\
4254.09 & 63: &$\pm$27\\
4275.53 & 28 &$\pm$4\\
4309.08 & 10 &$\pm$1\\
\end{tabular}
\end{minipage}
&
\begin{minipage}[!t]{26mm}
\begin{tabular}{lll}
\multicolumn{3}{l}{O{\sc ii}~(contd.)}\\
4317.14 & 22 &$\pm$5\\
4319.63 & 10: &$\pm$2\\
4327.47 & 10 &$\pm$1\\
4342.01 & 10: &$\pm$28\\
4349.43 & 25: &$\pm$2\\
4351.26 & 28 &$\pm$5\\
4366.89 & 24 &$\pm$3\\
4369.27 & 10 &$\pm$3\\
4395.94 & 11 &$\pm$2\\
4414.90 & 82: &$\pm$36\\
4416.97 & 30 &$\pm$7\\
4448.19 & 15 &$\pm$1\\
4452.38 & 13 &$\pm$1\\
4465.44 & 5 &$\pm$1\\
4488.19 & 5: &$\pm$1\\
4590.97 & 28 &$\pm$7\\
4596.18 & 47 &$\pm$7\\
4609.37 & 25 &$\pm$2\\
4638.85 & 63: &$\pm$27\\
4641.81 & 67 &$\pm$27\\
4661.63 & 21 &$\pm$4\\
4676.23 & 22 &$\pm$2\\
4699.19 & 31 &$\pm$4\\
4701.18 & 8 &$\pm$1\\
4703.16 & 8 &$\pm$1\\
4705.35 & 30 &$\pm$3\\
4710.01 & 13 &$\pm$4\\
4871.52 & 7 &$\pm$1\\
4906.83 & 9 &$\pm$1\\
4943.00 & 21 &$\pm$5\\[1mm]
Pb{\sc iv}\\
4496.15 & 51 &$\pm$6\\
\end{tabular}
\end{minipage}
\end{tabular}

\end{table}

\label{lastpage}
\end{document}